\documentclass[article]{lukaz}

\usepackage{lmodern}
\usepackage{framed}
\usepackage{booktabs}
\usepackage{multirow}
\usepackage[authoryear,round]{natbib}

\usepackage{cite}
\usepackage{amsmath,amssymb,amsfonts,algorithm, cancel}
\usepackage{framed}
\usepackage{graphicx}
\usepackage{textcomp}

\def\BibTeX{{\rm B\kern-.05em{\sc i\kern-.025em b}\kern-.08em
    T\kern-.1667em\lower.7ex\hbox{E}\kern-.125emX}}
    
\usepackage[noend]{algpseudocode}

\usepackage{graphicx}
\usepackage{amsmath,mathtools,amsthm}
\usepackage{amssymb}
\usepackage{amsbsy}
\usepackage{amsfonts}
\usepackage{algorithm, algorithmicx, algpseudocode}
\usepackage[justification=justified,singlelinecheck=false]{caption}
\usepackage[hang,flushmargin]{footmisc} 

\usepackage{listings}
\usepackage{pythonhighlight}

\usepackage{xcolor}
\definecolor{darkblue}{RGB}{0,0,128}
\definecolor{darkred}{RGB}{128,0,0}
\definecolor{darkgreen}{RGB}{0,128,0}
\newcommand{\red}[1]{{\color{red}#1}}
\newcommand{\blue}[1]{{\color{blue}#1}}
\newcommand{\darkgreen}[1]{{\color{darkgreen}\bf#1}}

\usepackage[export]{adjustbox}
\usepackage{multirow}
\usepackage{rotating}
\usepackage{booktabs}
\usepackage{subcaption}
\usepackage[utf8]{inputenc}

\newcommand{\x}{\mathbf{X}}
\newcommand{\y}{\mathbf{Y}}
\newcommand{\z}{\mathbf{Z}}

\newcommand{\B}{\mathcal B}
\newcommand{\C}{\mathcal C}

\newcommand{\R}{\mathbb R}
\renewcommand{\S}{\mathcal S}

\newcommand{\X}{\mathcal X}

\DeclareMathOperator*{\argmax}{\arg\!\max}

\newcommand{\N}{\mathbb N}

\newcommand{\as}{\text{~a.s.}}

\def\numdist{m}

\def\argmax{\operatorname{argmax}}
\def\polylog{\operatorname{polylog}}
\newcommand{\gap}{\lambda}
\newcommand{\mingap}{\lambda_{\min}}
\newcommand{\argmaxdisp}[1]{\underset{#1}{\argmax~}}

\author{Azadeh Khaleghi\\Mathematics \& Statistics\\
Lancaster University \And Lukas Zierahn\\  Computer Science\\
  Universit\`a degli Studi di Milano}
\title{\pkg{PyChEst}: a Python package for the consistent retrospective estimation of distributional changes in  piece-wise stationary time series } 

\begin{document}
\begin{abstract}
We introduce \pkg{PyChEst}, a \proglang{Python} package which provides tools for the simultaneous estimation of multiple changepoints in the distribution of piece-wise stationary time series. The nonparametric algorithms implemented are provably consistent in a general framework: when the samples are generated by unknown piece-wise stationary processes. In this setting, samples may have long-range dependencies of arbitrary form and the finite-dimensional marginals of any (unknown) fixed size before and after the changepoints may be the same. 
The strength of the algorithms included in the package is in their ability to consistently detect the changes without imposing any assumptions beyond stationarity on the underlying process distributions.  We illustrate this distinguishing feature by comparing the performance of the package against state-of-the-art models designed for a setting where the samples are independently and identically distributed. 
\end{abstract}
   \vskip 0.1in minus 0.05in
\section{Introduction} \label{sec:intro}
Retrospective multiple changepoint estimation is a classical problem in statistics, with 
a vast literature concerning both parametric  
and nonparametric settings, \citep{basseville:93,brodsky:93,muller1994change,brodsky2000non, Csorgo:98} .
The problem can be introduced as follows. 
A given sample
$$
\x=X_{1},\dots,X_{\tau_1}, 
X_{\tau_1+1},\dots,X_{\tau_2}, \dots,
X_{\tau_{\kappa}+1},\dots,X_n
$$ of length $n \in \N$
is formed as the concatenation of  $\kappa+1$  non-overlapping segments, 
where $\kappa \in 1, \dots, n$ and  $0<\tau_1 <\dots<\tau_\kappa<1$.  
Each segment is generated by some unknown stochastic process. 
The  process distributions of every pair of consecutive segments 
are different.  The index $\tau_k,~k \in 1,\dots,\kappa $ where   one segment ends and another starts is called a {\em changepoint}.
The changepoints $\tau_k$ as well as the number of changes $\kappa$ are  unknown and the objective is to estimate them given the sample $\x$. 

The changepoint estimation problem is usually considered in a  framework where the samples 
in each segment $X_{\tau_{k}+1}, \dots, X_{\tau_{k+1}}$ 
are assumed to be independently and identically distributed (i.i.d.) and the changes are
in the {\em mean} or in some fixed known moment, \citep{Yao:88,Carlstein:88, pan2006application, Lebarbier2005717,harchaoui2010multiple, ciuperca2011general, chen2011consistent, fryzlewicz2014wild}. 
In the literature concerning methods for dependent time series
the form of the change and the nature of 
dependence are usually restricted; for instance, 
the process distributions are assumed to have finite-memory 
or satisfy strong mixing conditions \citep{kokoszka1998change,  lavielle:99, kokoszka2002detection,davis2006structural, HarizWylieZhang2007, Lavielle:07, Brodsky:08, brodsky2013asymptotically}. 
Even if no restrictions are imposed on the memory of the processes, the finite-dimensional marginals before and after the changes are almost exclusively assumed to be different 
\citep{Carlstein:93,Giraitis:95}. 
On the other hand, existing changepoint estimation packages are typically concerned with independent observations,  where occasionally additional constraints such as parametric assumptions and moment conditions are placed on the distributions. In some cases the  included methods are designed to detect changes in the mean or variance only, and the settings considered do not allow for consistency considerations of the estimators \citep{JSSv058i03, JSSv066i03, JSSv062i07, truong2020selective}

We have implemented two consistent multiple changepoint estimators, proposed by \citet{khaleghi2012locating, khaleghi2014asymptotically}, which are designed to locate changes in the process distributions of highly dependent time series.  The only assumption made is that  each segment $X_{\tau_{k}+1}, \dots, X_{\tau_{k+1}}$  is generated by an 
(unknown) stationary ergodic  process. In particular, no such assumptions as independence, finite memory or mixing are placed on the samples. Moreover, the marginal distributions of any fixed size before and after the change may be the same:
the changes refer to those in the time-series process distributions. 
For example the mean, variance etc.\ may remain unchanged throughout the entire sequence. 
This means that the most general and perhaps the most natural notion of change is considered, 
that is, change in the distribution of the data. 
We also do not place
any conditions on the densities of the marginal distributions (the densities may not exist).
Since little assumptions are made on how the data are generated, 
this setting is especially suitable for highly dependent time-series, 
potentially accommodating a variety of new applications. 
To the best of our knowledge, 
\pkg{PyChEst} is the first software package to provide tools for the consistent estimation of distributional changes in time series with long-range dependencies. 

The remainder of this paper is organized as follows. In Section~\ref{sec:prob} we give a mathematical formulation of the changepoint problem  considered, and provide an overview of the methodology, as well as the possibilities and limitations of the framework that we consider. In Section~\ref{sec:tools} we introduce the \pkg{PyChEst} package, and describe how it can be installed and used,  and provide some examples and use-cases of the package.  In Section~\ref{sec:empiricial_verification} we illustrate the robustness  of our algorithms against long-range dependencies.

\section[]{Methodology}\label{sec:prob}
In this section we briefly introduce the setting considered, and provide an overview of the algorithms included in \pkg{PyChEst}, as well as a discussion of our implementation.
We refer to \citep{khaleghi2012locating, khaleghi2014asymptotically} for  detailed arguments and technical proofs. 

\subsection{Problem formulation} In order to formalize the problem, we require some preliminary notation and basic definitions. Let $(\X, \B_{\X})$ be a measurable space corresponding to the alphabet. 
Here, $\X$ could be a subset of $\R$ or more generally a subset of $\R^{d},~d\in \N$ equipped with its Borel $\sigma$-algebra. 
Let $\X^{\N}$ be the set of all $\X$-valued infinite sequences equipped with the Borel $\sigma$-algebra
$\B$ on $\X^\N$ generated by the cylinder sets. Stochastic processes are probability measures on $(\X^{\N},\B)$. 
Associated with a process $\mu$ is a sequence of random variables ${\bf X}:=\left ( X_t \right ),~ {t \in \N}$ with joint (process) distribution $\mu$ 
where for every $t \in \N$, $X_t:\X^\N \to \X$ is the coordinate projection of an in element of $\X^\N$ onto its $t^{\text{th}}$ element. 
We call a process $\mu$ stationary if
$\mu(X_{1}, \dots, X_{u}\in B)= \mu(X_{{1+j}}, \dots, X_{u+j} \in B)$ 
for all $B \in \B^{(u)},~u , j \in \N$.
For each $n \in \N$ and $B \in \B^{(u)},~u \in \N$ define the empirical measure
$\mu_n({\bf X},B): \X^\N  \rightarrow [0,1],~n \in \N$ of $B$ as
\[\mu_n({\bf X},B) :=
\frac{1}{n-u+1}\sum_{i=1}^{n-u+1} \mathbb{I}\{X_{i}, \dots, X_{i+u}\in B\}\] for $n \geq u$ and $0$ otherwise, where $\mathbb{I}$ is the indicator function.  
A stationary process $\mu$ with corresponding sequence of random variables {\bf X} is (stationary) ergodic if for every $u \in \N$ and $B \in \B^{(u)}$ it holds that 
$\lim_{n \rightarrow \infty} \mu_n({\bf X},B)=\mu(B),~\mu-\as$
This is equivalent to the standard definition involving the triviality of invariant measurable sets, see e.g. \citep{gray2009probability}.

The changepoint estimation problem considered in this paper can be formulated as follows. 
We are given a {\em piece-wise stationary} an $\X$-valued sample $\x:=X_1,\dots,X_n$ of length $n\in \N$, 
formed as the concatenation of a number $\kappa+1$ of 
stationary segments, 
\[
X_{1},\dots, X_{\tau_1},
X_{\tau_1+1}, \dots, X_{\tau_2}, \dots,
X_{\tau_{\kappa}+1}, \dots,X_{n}
\]
with {\em changepoints} at $\tau_1<\dots<\tau_{\kappa}$ where 
each (stationary) segment is generated 
by one of $\numdist \leq \kappa+1$ unknown {\em stationary ergodic} processes $\mu_1,\dots,\mu_{\numdist}$. In Section~\ref{sec:prob} we described the changepoint problem, and informally stated that a piece-wise stationary sample $\x$ is obtained as a concatenation of $\kappa+1$ segments each generate by a stationary ergodic process.  More formally, the process by which $\x$ is obtained can be specified as follows.
First, let $\rho$ 
be a measure on the space $(\X^{{(\kappa+1)} \times \N}, \B^{\otimes {(\kappa+1)}})$ where,
$\B^{\otimes {(\kappa+1)}}:= \sigma(\{B_1 \times \dots \times B_{\kappa}: B_k \in \B,~k \in 1, \dots, \kappa\}).$
Define the infinite matrix of $\X$-valued random variables ${\mathbb X} :=
    \begin{pmatrix}
    X_{i,j}
    \end{pmatrix}, {i \in 1, \dots, \kappa+1, j \in \N}$
where $X_{i,j}:\X^{{(\kappa+1)} \times \N} \rightarrow \X,~i \in 1,\dots, \kappa+1,j \in \N$ are jointly distributed according to ${\rho}$, so that for $B \in \B^{\otimes {(\kappa+1)}}$ we have $\Pr({\mathbb X} \in B) = \rho(B)$. For each $i \in 1, \dots, \kappa+1$, let $\x_i:= \left ( X_{i,j} \right), j \in \N$ and define the projection map 
$\pi_i \mapsto \x_i$ which maps $\mathbb X$ on its $i^{\text{th}}$ row. The marginal distribution $\mu_i$ of $\x_i$ is then defined as the distribution induced by $\rho$ on the $i^\text{th}$ row of $\mathbb X$, i.e. 
$\mu_i:=\rho \circ \pi_i^{-1}$. Next, define the mapping $c: \N \rightarrow \N \times \N$ as 
$c(j)\mapsto (t^*(j)+1,j-\tau_{t^*(j)})$
where  $t^*(j):=\max_{i \in 0,\dots,\kappa+1} \tau_i \leq j$ picks out the changepoint $\tau_i$ that is closest to $j \in \N$ from the left, with the convention that $\tau_0:=0$ and $\tau_{\kappa+1}:=n$.
The {\em piecewise stationary sample} $\x$ generated by $\rho$ with changepoints at $\tau_1,\dots,\tau_{\kappa}$ can be specified as a sequence of coordinate projections $X_t: \X^{n} \rightarrow \X,~t \in 1, \dots, n$ such that for any $\ell \in 1,\dots, n,~t_1, \dots, t_\ell \in 1, \dots, n$ and $B_i \in \B_{\X},~i\in 1, \dots, \ell$ it holds that
\begin{equation*}
\Pr(X_{t_1} \in B_1, \dots,X_{t_\ell} \in B_\ell )= \rho(X_{c(t_1)} \in B_1, \dots,X_{c(t_\ell)} \in B_\ell).
\end{equation*}
It is straightforward to check that a set-function satisfying the above, extends to a probability measure on $(\X^n,\B^{(n)})$, with the property that  the (marginal) distributions of the segments $X_{\tau_k+1},\dots,X_{\tau_{k+1}}$ are each given by a stationary ergodic process distribution $\mu_{k},~k \in 1,\dots,\kappa+1$. Since it is assumed that $\mu_k \neq \mu_{k+1},~i \in 1,\dots, \kappa$, the indices $ \tau_k,~k \in 1,\dots, \kappa$ are called {\em changepoints}. 
The consecutive segments are generated by {\em different} processes so that $\mu_{k} \neq \mu_{k+1},~k \in 1,\dots,\kappa$.
However, their means, variances, or, more generally, their finite-dimensional
marginal distributions may be the same upto some (unknown) fixed size. 
On the other hand, letting $\numdist\leq \kappa+1$ means that non-consecutive segments can have the same distribution. 
Observe that by this formulation, the finite-dimensional marginals of any fixed size
before and after the changepoints $\tau_k$ may be the same. 
We consider a general notion, where the changes are in the 
{\em process distributions}. 
Our aim is to provide {\em  asymptotically consistent} estimates $\widehat{\tau}_k(n)$ of the unknown changepoints $\tau_k,~k \in 1,\dots,\kappa$. 
An estimate  $\widehat{\tau}_k(n)$ is said to be {\em asymptotically consistent} if
\begin{equation}\label{eq:const}
\lim_{n\rightarrow \infty}  \frac{1}{n} |\widehat{\tau}_k(n)-\tau_k| = 0, \as.
\end{equation}
We seek changepoint estimation algorithms that provide an asymptotically consistent estimate 
for every changepoint $\tau_k,~k=1,\dots,\kappa$. 
Observe that the asymptotic regime simply means that 
the estimation error is arbitrarily small if the sequence is sufficiently long.
That is, the problem is to retrospectively locate the changes and the given sample $\x$ does not grow with time.  
This differs, for example, from the problem of online changepoint detection where the observations arrive sequentially over time, and the objective is to {\em detect} 
a change as soon as possible. 
Denote by 
\[
\gap(n):=\min_{k=1, \dots, \kappa+1} \frac{1}{n} (\tau_k-\tau_{k-1})
\]
with $~k \in 1,\dots,\kappa+1,~\tau_0:=0,~\tau_{\kappa+1}:=n$, 
the minimum normalized length of the stationary segments, and let 
\[\mingap:=\liminf_{n\rightarrow \infty}\gap(n).\] 
In order for the consistency conditions of the algorithms implemented as part of this package to hold we require that $\mingap>0$. 
Note that this linearity condition is standard in the changepoint literature, 
although it may be unnecessary when simpler formulations of the problem 
are considered.  Indeed, an equivalent assumption is made by \citet[Assumption 3.2,  3.3]{fryzlewicz2014wild} for the simpler piece-wise i.i.d.  setting. 
However, conditions of this kind are  
 inevitable in the general setting that we consider,
where  there may be arbitrary long-range dependencies:
 if the length of one of the segments is constant or 
sub-linear in $n$ then asymptotic consistency is not possible in this setting.  
\subsection{A distance between process distributions}  Most non-parametric approaches to estimating a changepoint  
are usually based on the following idea. 
Initially, every possible index $i \in \{1,\dots,n\}$ is considered a potential changepoint. 
The difference between the empirical expectation 
of the two segments $X_1,\dots, X_i$ and $X_{i+1}, \dots, X_n$ 
on either side of every fixed $i \in  \{1,\dots,n\}$ is calculated, and the changepoint estimate 
is chosen as the index that maximizes this difference in absolute value. 
A more general approach is based on 
maximizing the difference between the empirical distributions 
of the two segments under a given norm. 
Different norms give rise to different test statistics. 
The commonly used distances include the Kolmogorov-Smirnov statistics, 
obtained when the difference is calculated under the $L_\infty$ norm, 
the Cram\'er-von Mises statistics corresponding to the use of $L_2$ norm, 
and the generalizations thereof.  
We rely on empirical estimates of a {\em distributional distance} between the underlying process measures to locate the changepoints. 
The distance $d(\mu_1,\mu_2)$ 
between a pair of process distributions $\mu_1, \mu_2$ can be defined as 
\[d(\mu_1,\mu_2):=\sum_{i \in \N} w_i |\mu_1(A_i)-\mu_2(A_i)|\]
where, $w_i$ are positive summable real weights, and $A_i$ range
over a countable field that generates the sigma-algebra of the underlying probability space.
For a discrete alphabet $A_i$ range over the set of all possible tuples. 
For example, in the case of binary alphabets, 
the distributional distance is the weighted sum of the differences of the probability values 
(measured with respect to $\mu_1$ and $\mu_2$)
of all possible tuples $0, 1, 00, 01, 10, 11, 000, 001,\dots$. 
For real-valued processes the sets  $A_i$ 
range over the products of all intervals with rational endpoints 
(i.e. the intervals, all pairs of such intervals, triples, etc.). 
Asymptotically consistent estimates of this distance 
can be obtained by replacing unknown probabilities with the corresponding frequencies, provided  
that the corresponding process distributions are stationary ergodic. 
Although the distance involves infinite summations, the computational complexity of obtaining its empirical approximation is $\mathcal O(n \polylog n)$, see  \citep{khaleghi2016consistent} for a discussion.  Our implementation of this distance relies on the Aho-Corasick Algorithm \citep{aho1975efficient} for calculating empirical frequencies. 

In principal, we can use any distance function between the samples, 
provided that the distance used reflects that between the underlying process distributions. 
In the case of changepoint analysis, the distance is required to satisfy convexity as well. 
It is important to note that the distinction between the underlying process distributions 
is not reflected 
by string metrics, such as the Hamming distance, or the Levenshtein distance, etc.
The Hamming distance 
between two sequences is defined as
the minimum number of substitutions required to transform one sequence into another, 
and the Levenshtein distance
corresponds to the smallest number of deletions, insertions
and substitutions needed to achieve the same objective, see e.g.\ \citep{stephen:94}.
More generally, a string distance between a pair of sequences is $0$ if and only if 
the two sequences are exactly the same. 
However, what we require is for the distance to converge to  
$0$ for long enough samples, if and only if both sequences 
are generated by the same process distribution. 
Consider a simple example where 
the elements of the sequences 
$\x:=X_1,\dots,X_n $ and $\y:=Y_1,\dots,Y_n$
are drawn independently 
from a Bernoulli distribution with probability $p:=1/2$. 
Regardless of the value of $n$, on average, 
the Hamming distance between the two sequences 
is $1/2$ while they are generated by the same process distribution. 
At the same time,  
the empirical estimate of the distributional distance between $\x$ and $\y$ 
becomes arbitrarily small for large enough $n$.  
\subsection{Detecting the number of changepoints} \label{sec:changepoints}
In general, rates of convergence are
provably impossible to obtain for stationary ergodic processes \citep{Sheilds:96}. As a result, the two-sample test does not have a consistent solution in this framework \citep{Ryabko:10discr}. This means that  it is not possible to determine, 
even in the weakest asymptotic sense, whether or not two samples have been generated by the same or by different stationary ergodic distributions. An implication of this negative result is that it is provably impossible to determine the number of changes $\kappa$ without making further assumptions. 
In light of this limitation, without placing any additional constraints on the underlying process distributions, in this software package we provide two options. 
With this theoretical impossibility result in mind,  \pkg{PyChEst} provides two (consistent) algorithms as follows. 
\begin{itemize}
\item The first algorithm is an implementation of the the so-called List-Estimator of \citet{khaleghi2012locating},
which aims to produce a (sorted) list of candidate estimates 
whose first $\kappa$ elements are consistent estimates of the true changepoints.  
More precisely, 
the algorithm generates an exhaustive list of possibly more than $\kappa$ 
candidate estimates (but makes no attempt to estimate $\kappa$). 
The produced list must have the property that 
its first $\kappa$ elements are consistent estimates of the true changepoints. 
In order to achieve this goal, the list-estimator requires an additional parameter $\alpha \in (0,1)$, 
which is a lower-bound on the minimum separation 
$\mingap$. It is worth noting that if the correct value of $\kappa$ is known a-priori, it is then possible to consistently estimate the changepoints 
without any additional information, and in particular access to some $\alpha \leq \mingap$ would no longer be required \citep{khaleghi2016nonparametric}.

\item The second algorithm included in the package,  is  an implementation of the procedure proposed by \citet{khaleghi2014asymptotically}, 
which is able to consistently estimate the changepoints and detect the number of changes, provided that
the total number $\numdist$ of different
process distributions is given. 
Note that in the specific case where all  of the process distributions are different, 
knowing $\numdist$ amounts to knowing the number 
of changepoints ($\numdist=\kappa+1$).
However, the required additional parameter $\numdist$ can be in general very different from $\kappa+1$, and
has  a natural interpretation in many real-world applications. 
For instance, 
the sequence $\x$ may correspond to the behavior of a system over time,
which may have alternated $\kappa>\numdist-1$ times between a known number $\numdist$ of states. 
In a simple case, the system may only take on $\numdist=2$ states, 
for example, ``normal'' and ``abnormal''. 
Another application of this setting would be in speech segmentation where the total number $\numdist$ of speakers is known, 
while the number $\kappa$ of alternations between the speakers is not available. 
Thus, the number $\numdist$ of process distributions is provided as an intrinsic part of the problem, 
which is provably sufficient to estimate $\kappa$. See Section~\ref{app:algos} for an informal overview of the algorithms as well as details concerning their implementation; more technical discussions can be found in the corresponding papers referenced above. 
\end{itemize}

\subsection{An overview of our implementations}\label{app:algos}
In  this section we give an overview of the changepoint estimation algorithms provided in the package focusing on our algorithmic implementations. 
Let us denote by $\widehat{d}(\cdot,\cdot)$ the empirical estimate of the distributional distance $d(\cdot,\cdot)$; 
we refer to \citep{khaleghi2016consistent} for a definition and a discussion on computational considerations. 

The following two operators namely, the 
intra-subsequence distance $\Delta_{\x}$ 
and the single-changepoint estimator $\Phi_{\x}$ 
are used in our methods.
Let $\x=X_{1},\dots,X_{n}$ be a sample generated by a stochastic process
and consider a subsequence $X_{a},\dots,X_{b}$ of $\x$ with $a < b \in 1,\dots,n$.  
\begin{enumerate}
\item[i.~] Define the intra-subsequence distance of $X_{a},\dots,X_{b}$ as
\begin{equation}\label{defn:Delta}
\Delta_{\x}(a,b):= \widehat{d}(X_{a},\dots, X_{ \lfloor \frac{a+b}{2}\rfloor},X_{\lceil \frac{a+b}{2}\rceil},\dots,X_{b} )
\end{equation}
\item[ii.~] For some $\alpha \in (0,1)$, define the single-changepoint-estimator of $X_{a},\dots,X_{b}$ as
\begin{equation}\label{defn:Phi}
\Phi_{\x}(a,b,\alpha):=~\argmaxdisp{t \in a,\dots,b}~\widehat{d}(X_{a-\lceil n\alpha\rceil},\dots,X_{t},X_{t+1},\dots,X_{b+\lfloor n\alpha\rfloor })
\end{equation}
\end{enumerate}
For simplicity of notation, from this point on we assume the floor and ceiling functions implicit.
\subsubsection{List-Estimator}
The changepoint estimation procedure proposed by \citet{khaleghi2012locating} works as follows. 
Given a piece-wise stationary sample $\x=X_{1},\dots, X_n$ and a lower-bound $\alpha \in (0,1)$ on the minimum (normalized) stationary segment length, 
a sequence of evenly-spaced indices is formed. 
The index-sequence partitions $\x$ into consecutive 
segments of length $\alpha/3$. 
The intra-subsequence-distance $\Delta_\x$ is calculated in all but the first and last segments and is 
used as their performance scores. 
Next, an appropriate subset of the segments is selected through the following iterative procedure:
initially, all but the first and last segment are marked as available, and 
at each iteration, an available segment of {\em highest score} is selected, and added to the final list. 
Noting that $\alpha$ is a lower-bound on the minimum normalized distance between the changepoints, 
the available candidates within $n\alpha/2$ of the selected segment
are made unavailable. The selection procedure continues until 
the set of available segments becomes empty. 
This is depicted below, where the downward arrows show the location of the indices. 
\begin{align*}
&\underbracket[2pt]{X_1,X_2,\dots~~}\overset{\darkgreen{\displaystyle \boldsymbol \downarrow}}{~} 
\overset{\text{\textcolor{red}{Segments initially made available}}}{\overbrace{{
\underset{\Delta_\x{~\text{as score}}}{\underbracket[2pt]{\textcolor{blue}{~~~\leftarrow \alpha n/3 \rightarrow~~}}}}\overset{\darkgreen{\displaystyle \boldsymbol \downarrow}}{~}
\underset{\dots}{\textcolor{white}{\dots}}\overset{\darkgreen{\displaystyle \boldsymbol \downarrow}}{~}
\underset{\Delta_\x{~\text{as score}}}{{\underbracket[2pt]{\textcolor{blue}{~~~\leftarrow  \alpha n/3 \rightarrow~~}}}}\overset{\darkgreen{\displaystyle \boldsymbol \downarrow}}{~} 
\underset{\dots}{\textcolor{white}{\dots}}
\overset{\darkgreen{\displaystyle \boldsymbol \downarrow}}{~}
{{\underset{\Delta_\x{~\text{as score}}}{\underbracket[2pt]{{\textcolor{blue}{~~~\leftarrow  \alpha n/3 \rightarrow~~}}}}}}}}
\overset{\darkgreen{\displaystyle \boldsymbol \downarrow}}{~}
{\underbracket[2pt]{\textcolor{white}{~~~~~~~}\dots,X_n~}}
\end{align*}
Noting that the first and last segments are made unavailable at the outset, it is easy to see that by 
this procedure, the segments to the left and right of each available segment are made unavailable. 
The single-changepoint estimator $\Phi_\x$ is used to generate 
a candidate estimate within every selected segment. 
There may be more than $\kappa$ candidate estimates produced;
however, as shown by \citet{khaleghi2012locating}
for long enough $\x$ 
the first $\kappa$ candidate estimates in the  sorted list 
are consistent. We would like to point out that since the candidate estimates can be computed in parallel, our implementation of this algorithm is multi-threaded. 

An informal argument as to why this procedure is consistent is as follows. When $\alpha \leq \mingap$, 
the generated index-sequence   partitions $\x$ in such a way that 
every three consecutive segments of the partition contain {\em at most} 
one changepoint. Also, the segments are of lengths $\alpha n/3$. 
Therefore, if $n$ is large enough, the single-changepoint-estimator $\Phi_\x$ produces 
consistent changepoint estimates within each of the   
segments that actually contains a true changepoint.
Moreover,  the performance scores of  
the segments without changepoints converge to $0$, while 
each of 
those corresponding to the segments that 
contain a changepoint converge to a non-zero constant.  
Thus, the $\kappa$ 
candidate estimates of 
highest performance score that are 
at least at a distance $\alpha n$ from one another,  
each consistently estimate a unique changepoint.
We refer to the above reference for a rigorous proof of consistency. 
\subsubsection{Changepoint Estimator} 
The function called Changepoint Estimator in our package is an implementation of the algorithm proposed by \citet{khaleghi2014asymptotically}. 
Given a piece-wise stationary sample $\x=X_{1},\dots, X_n$, a lower-bound $\alpha \in (0,1)$ on the minimum (normalized) stationary segment length, and the total number of different process distributions $\numdist$, the algorithm works as follows. 
First, the List-Estimator described is used to obtain an initial set of $K \leq \alpha^{-1}$ changepoint candidates. 
The estimates are sorted in {\em increasing order appearance} to give $\widehat{\tau}_1,\dots,\widehat{\tau}_K$. The sorted changepoint estimates are in turn used to produce a set 
\[\S:=\{\x_i:=X_{\widehat{\tau}_{i-1}}, \dots, X_{\widehat{\tau}_i}:i \in 1,\dots,K+1,~\widehat{\tau}_{0}:=1,~\widehat{\tau}_{K}:=n \}\]
of at most $\lfloor \alpha^{-1} \rfloor+1$ consecutive non-overlapping segments of $\x$, which are of possibly different lengths. 
The identified segments of $\x$ are depicted below.
\begin{align*}
&\underset{\x_1}{\underbracket[2pt]{\overset{X_1,X_2,X_3,\dots~~~~~~~}{~~~~~~~~~~~~~~}}}\overset{\blue{\overset{\widehat{\tau}_1}{\boldsymbol \star}}}{}
\underset{\x_2}{\underbracket[2pt]{~~~~~~~~}}\overset{\blue{\overset{\widehat{\tau}_2}{\boldsymbol \star}}}{} 
\underset{\dots}{\textcolor{white}{\quad \dots \quad}}\overset{\blue{\overset{\widehat{\tau}_{i-1}}{\boldsymbol \star}}}{}
\underset{\textcolor{white}{~\leftarrow~\geq \alpha n~\rightarrow}}{
\underset{\x_i}{\underbracket[2pt]{~~~~~~~~~~~~}}}\overset{\blue{\overset{\widehat{\tau}_{i}}{\boldsymbol \star}}}{}
\underset{\dots}{\textcolor{white}{\quad \dots \quad}}
\overset{\blue{\overset{\textcolor{blue}{\widehat{\tau}_{K-1}}}{\boldsymbol \star}}}{}
\underset{\x_{K}}{\underbracket[2pt]{~~~~~~~~~~~~~~~~~~}}\overset{\blue{\overset{\widehat{\tau}_{K}}{\boldsymbol \star}}}{} 
\underbracket[2pt]{\overset{~~~\dots, X_n}{~~~~~~~~~}}
\end{align*}
Next, a procedure based on time-series clustering is employed to remove redundant estimates. This is where the total number $\numdist$ of process distributions is used. 
To this end, first  a total of $\numdist$ {\em cluster centers} are obtained as follows.
The first segment $\x_1$ is chosen as the first cluster center. 
Iterating over $j=2, \dots, \numdist$ a segment is 
chosen as a cluster center if it has the highest minimum distance from 
the previously chosen cluster centers. 
Once the $\numdist$ cluster centers are specified, the remaining 
segments are assigned to the closest cluster. As shown by \citet{khaleghi2014asymptotically}, 
for a long enough $\x$, this clustering algorithm will put the segments together if and only if they have the same (unknown) process distribution. 
In each cluster, the changepoint candidate
that joins a pair of consecutive segments of $\x$ is naturally identified as {\em redundant} 
and is removed from the list. 
This step of the algorithm is pictured below where $\C_i$ denotes the $i^{\text{th}}$ cluster, $i \in 1,\dots,\numdist$. 
\begin{align*}
&
\underset{\dots}{\textcolor{white}{\dots}}
\textcolor{blue}{\underset{\in \C_1}{\underbracket[2pt]{~~~~~~~~~~~~~~}}}\overset{{\overset{\xcancel{{\widehat{\tau}_{j-1}}}}{{\boldsymbol \star}}}}{}
\textcolor{blue}{\underset{\in \C_1}{\underbracket[2pt]{~~~~~~~~}}}
\overset{\red{\overset{\widehat{\tau}_j}{\boldsymbol \star}}}{}
\darkgreen{\underset{\in \C_2}{\underbracket[2pt]{~~~~~~~~~~~~}}}\overset{{\overset{\xcancel{{\widehat{\tau}_{j+1}}}}{{\boldsymbol \star}}}}{}
\darkgreen{\underset{\in \C_2}{\underbracket[2pt]{~~~~~~~~~~~~~~~~~~}}}\overset{\red{\overset{{{\widehat{\tau}_{j+2}}}}{{\boldsymbol \star}}}}{} 
\blue{\underset{\in \C_1}{\underbracket[2pt]{~~~~~~~~~}}}
\underset{\dots}{\textcolor{white}{\dots}}
\end{align*}
In the above example, $\widehat{\tau}_{j-1}$ and $\widehat{\tau}_{j+1}$ each separate segments that, as identified by the clustering sub-routine, are generate by the same unknown process, and are thus considered as redundant. Once all of the redundant candidates are removed, 
the algorithm outputs the remaining candidate estimates. 

An intuitive argument concerning the consistency of this procedure is as follows.  
First, note that for a long enough $\x$ the List-Estimator produces
an initial set of at least $\kappa$ 
estimates whose first $\kappa$ elements each uniquely and consistently estimate one of the $\kappa$ changepoints. 
(Since $\kappa$ is unknown, the fact that the correct estimates are located first in the list is irrelevant: what matters is that the correct changepoint estimates are somewhere in the list.) 
Therefore, if $\x$ is long enough, the largest portion of each segment in 
$\S$ is generated by a single process distribution.
Since the List-Estimator ensures that the initial changepoint candidates 
are at least $n\alpha$ apart,
the lengths of the segments in $\S$ are linear in $n$. 
Thus, it can be shown that for large enough $n$,  the empirical estimate of the distributional
distance between a pair of segments in $\S$ converges 
to $0$ if and only if the same process distribution generates {\em most of} the two segments. 
Given the total number $\numdist$ of process distributions, 
for long enough $\x$  the clustering algorithm groups together those and only those segments in $\S$ 
that are generated by the same process.  
This lets the algorithm identify and remove the redundant candidates. 
As a result of the consistency of the List-Estimator, the remaining candidates are consistent estimates of the true changepoints.
The proof can be found in \citep{khaleghi2014asymptotically}.

\section[]{Installation and Use}\label{sec:tools}
In this section we outline how to install and use \pkg{PyChEst}. 

\subsection[]{Installing \pkg{PyChEst}}
The easiest way to install the \code{Python} package is to use \code{Python}s package-management system \code{pip}, which can be achieved by typing the following segment into a command line of their choice:
\begin{CodeInput}
pip install PyChest
\end{CodeInput}

\subsection[]{Using \pkg{PyChEst}}
The \pkg{PyChEst} package provides two main functions,  namely
\begin{enumerate}
\item \code{list\_estimator} 
\item \code{find\_changepoints} 
\end{enumerate} 
both described in the previous section.  The \code{list\_estimator} takes a sample and a parameter \code{min_distance} corresponding to a lower bound on the minimum normalized distance $\lambda_{\min}$ between the changepoints. Note that this parameter is denoted by $\alpha$ in the mathematical descriptions given in  Sections~\ref{sec:changepoints}~and~\ref{app:algos}.  The function can be called as follows.
\begin{CodeInput}
>>> estimates = PyChest.list_estimator(seq, min_distance)
\end{CodeInput}
The function \code{find\_changepoints} requires the same parameters as well as an additional integer corresponding to the total number of processes that generate the data; it can be called as described below.
\begin{CodeInput}
>>> estimates = PyChest.find_changepoints(seq, min_distance, process_count)
\end{CodeInput}
Next,  we demonstrate how the package can be used. To this end, first we load the required packages as follows.
\begin{CodeInput}
>>> import PyChest
>>> import numpy as np
\end{CodeInput}
We also  fix a random seed for reproducibility.
\begin{CodeInput}
>>> np.random.seed(1)
\end{CodeInput}
We generate a piece-wise i.i.d.  Bernoulli sample with changepoints at 2000 and 6500,  where the parameter corresponding to the first and last segments is $0.2$ and that of the middle segment $0.7$, i.e.
\begin{CodeInput}
>>> seq = []
>>> seq.append(np.random.binomial(1, p=0.2, size=2000))
>>> seq.append(np.random.binomial(1, p=0.7, size=4500))
>>> seq.append(np.random.binomial(1, p=0.2, size=1500))
>>> seq = np.concatenate(sequence)
\end{CodeInput}
Notice that the data so-generated is a piece-wise i.i.d.  sample with changepoints at $2000$ and 
$6500$,  with the minimum length of the stationary segment equal to $1500$.  There are a total two processes present,  since the first and last segments have been generated by the same process.  We can now execute \pkg{PyChest} to find the changepoints.  Both the \code{list\_estimator} and \code{find\_changepoints} a lower bound on the minimum normalized distance $\lambda_{\min}$ between changepoint.  The shortest stationary segment in this example is of length $1500$,  and given that the sequence is of length $8000$,  the minimum normalized stationary segment length equals $\lambda_{\min}=0.1875$.   We assume this value unknown,  and arbitrarily provide a parameter $\alpha=0.125 \leq \lambda_{\min}$ as input.
\begin{CodeInput}
>>> min_distance = 0.125
>>> estimates = PyChest.list_estimator(seq, min_distance)
\end{CodeInput}
The \code{list\_estimator}  returns a \code{Python}  of changepoint estimates at least $0.125\times 8000$ apart,  which are ordered in decreasing sequence of their {\em performance scores} $\Delta_{\x}$ as given by \eqref{defn:Delta}.  
As previously discussed,  the theoretical consistency results guarantee that for a long enough sequence,   with probability one,  the first $2$ estimates in the list are arbitrarily close to the true changepoints.   
\begin{CodeInput}
>>> print(estimates)
\end{CodeInput}
\begin{CodeOutput}
[1997, 6502, 4572] 
\end{CodeOutput}
As can be seen above,  the first two elements of the array returned by  \code{list\_estimator}  are good estimates of the true changepoints $\tau_1=2000$ and $\tau_2=6500$ in our example.

In addition to \code{seq} and \code{min\_distance},  the \code{find\_changepoints} function also requires the total number of different processes that generate the data.  Since there are three segments in our example with the first and last segments  generated by the same process,  the total number of processes here is $2$.  We refer to this parameter as \code{process_count} and note that it has been mathematically denoted by $m$ in Section~\ref{app:algos}.
\begin{CodeInput}
>>> process_count = 2
>>> estimates = PyChest.find_changepoints(seq, min_distance, process_count)
\end{CodeInput}
\begin{CodeInput}
>>> print(estimates)
\end{CodeInput}
\begin{CodeOutput}
[1997, 6502] 
\end{CodeOutput}

Next,  we generate a real-valued sequence from the above binary-valued sequence by calculating its running mean (of window size $N=25$) as follows.  
\begin{CodeInput}
>>> N = 25
>>> seq = np.convolve(seq, np.ones(N)/N, mode='valid')
\end{CodeInput}
Applying \code{list_estimator} and  \code{find_changepoints} to the generated data we obtain the following results. 
\begin{CodeInput}
>>> min_distance = 0.125
>>> estimates = PyChest.list_estimator(seq, min_distance)
\end{CodeInput}
\begin{CodeInput}
>>> print(estimates)
\end{CodeInput}
\begin{CodeOutput}
[1989, 6489, 4582]
\end{CodeOutput}

\begin{figure}
\begin{center}
\includegraphics[scale=1]{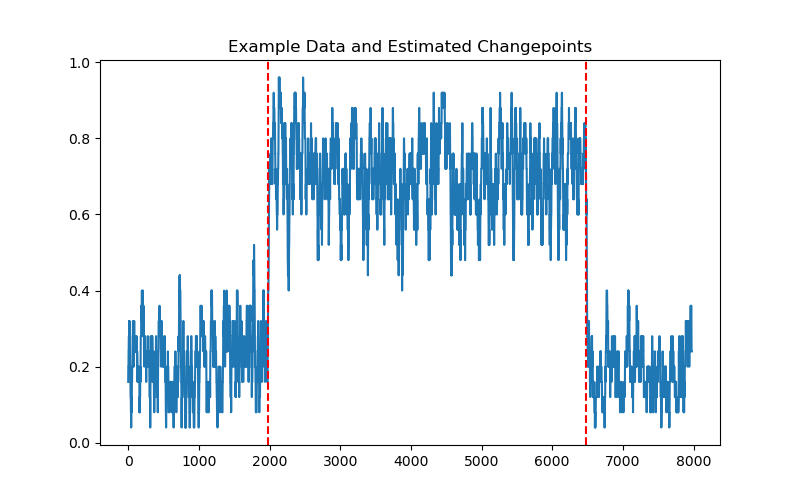}
\caption{An illustration of the changepoint estimates  produced by \code{find\_changepoints}.}\label{fig:conv}
\end{center}
\end{figure}

\begin{CodeInput}
>>> process_count = 2
>>> estimates = PyChest.find_changepoints(seq, min_distance, process_count)
\end{CodeInput}
\begin{CodeInput}
>>> print(estimates)
\end{CodeInput}
\begin{CodeOutput}
[1989, 6489] 
\end{CodeOutput}
The result is visualized in Figure~\ref{fig:conv}.
Observe that while the illustration in this example refers to the estimation of changes in the mean,  the algorithms are more generally designed to detect changes in the distribution.  This is shown in the next section.

\section[]{Robustness against long-range dependencies} \label{sec:empiricial_verification}
The strength of \pkg{PyChEst} is in its ability to consistently estimate changepoints in the presence of long-range dependencies. 
In this section we showcase this feature by comparing \pkg{PyChEst} to the state of the art algorithm Wild Binary Segmentation (WBS) proposed by \citet{fryzlewicz2014wild} which is provably consistent in the piece-wise i.i.d.  setting; we rely on the implementation provided by the CRAN package \pkg{wbs} of \citet{wbs-CRAN}.  

\subsection[]{Stationary samples which possess long-range dependencies}
As part of the experiments outlined in Section~\ref{sec:illus} we obtain the stationary segments of some of our piece-wise stationary samples as sample-paths of a subclass of stationary ergodic processes that do not belong to any ``simpler'' class. 
More specifically,  we consider Irrational Rotations outlined below,  which  
are classical examples of a stationary ergodic process that is not a $B$-process, and that cannot 
be modelled by a hidden Markov model with a finite or countably
infinite set of states, see, e.g. \citep{Sheilds:96}.  
\subsubsection{Irrational Rotation} For a fixed $t \in \N$,  a binary-valued  sample $\y := Y_1,Y_2,\dots,Y_t$ is generated from an Irrational Rotation as follows.  First, we fix a parameter  $\beta\in(0,1)$.
Then,  we sample  $R_0$ uniformly at random from $[0,1]$, and for each $i=1,\dots, t$ we obtain $R_i$ by shifting
$R_{i-1}$ by $\beta$ to the right and removing the integer part, i.e. 
\begin{equation*}
R_i := R_{i-1}+ \beta -\lfloor R_{i-1}+ \beta \rfloor,~i = 1,\dots,t
\end{equation*}
The sequence $\y$ is obtained from $R_i, ~i \in 1,\dots, t$ by thresholding at $0.5$ i.e., for each $i \in 1, \dots, t$ we set
\begin{equation}\label{eq:bir}
Y_i:=\mathbb{I}\{R_i \geq 0.5\}
\end{equation}
It is well-known that if $\beta$ is irrational then $Y$
forms a binary-valued stationary ergodic sample,  that possesses long-range dependencies: it is not even  mixing in the ergodic-theoretic sense \citep{Sheilds:96}.
We simulate $\beta$ by a floating point number with a long mantissa.
\subsubsection{Hidden Irrational Rotation}
In order to obtain a sample $\z:=Z_1,\dots,Z_t,~t \in \N$ from a real-valued stationary ergodic process that is not mixing,  we generate what we call a ``Hidden Irrational Rotation''.  To this end,  we first simulate an irrational number $\beta$ by a floating point number with a long mantissa,  and generate a binary-valued Irrational Rotation sample $\y:=Y_1,\dots,Y_t$ given by \eqref{eq:bir}.  Next, we draw two independent samples $U_1, \dots, U_t$ and $V_1 \dots, V_t$,  each generated uniformly i.i.d.  from $[0, 1]$ and $[0.9, 1.9]$ respectively.  We obtain $\z$ as
\begin{equation}
Z_i:= U_i (1-Y_i) + V_iY_i,~i=1,\dots,t
\end{equation} 
This forms a real-valued  (non-mixing) stationary ergodic sample $\z$. 
Observe that if we deterministically set $U_i=0$ and $V_i=1$ for all $i=1,\dots,t$ we recover the binary-valued sample $\y$.  
Figure~\ref{fig:hir} illustrates an example of a piece-wise stationary sample,  with a changepoint at $\tau=1500$ such that the segments before and after it correspond to  sample-paths of two different Hidden Irrational Rotations with parameters $0.45..$\footnote{$0.452341643253462432$} and $0.63..$\footnote{$0.6345354645623456234234$} respectively.  Observe that while there is a change in the process distributions,  it cannot be located visually and without appropriate statistical tools.
\subsection{Illustrations}\label{sec:illus}
\begin{figure}
\begin{center}
\includegraphics[scale=1]{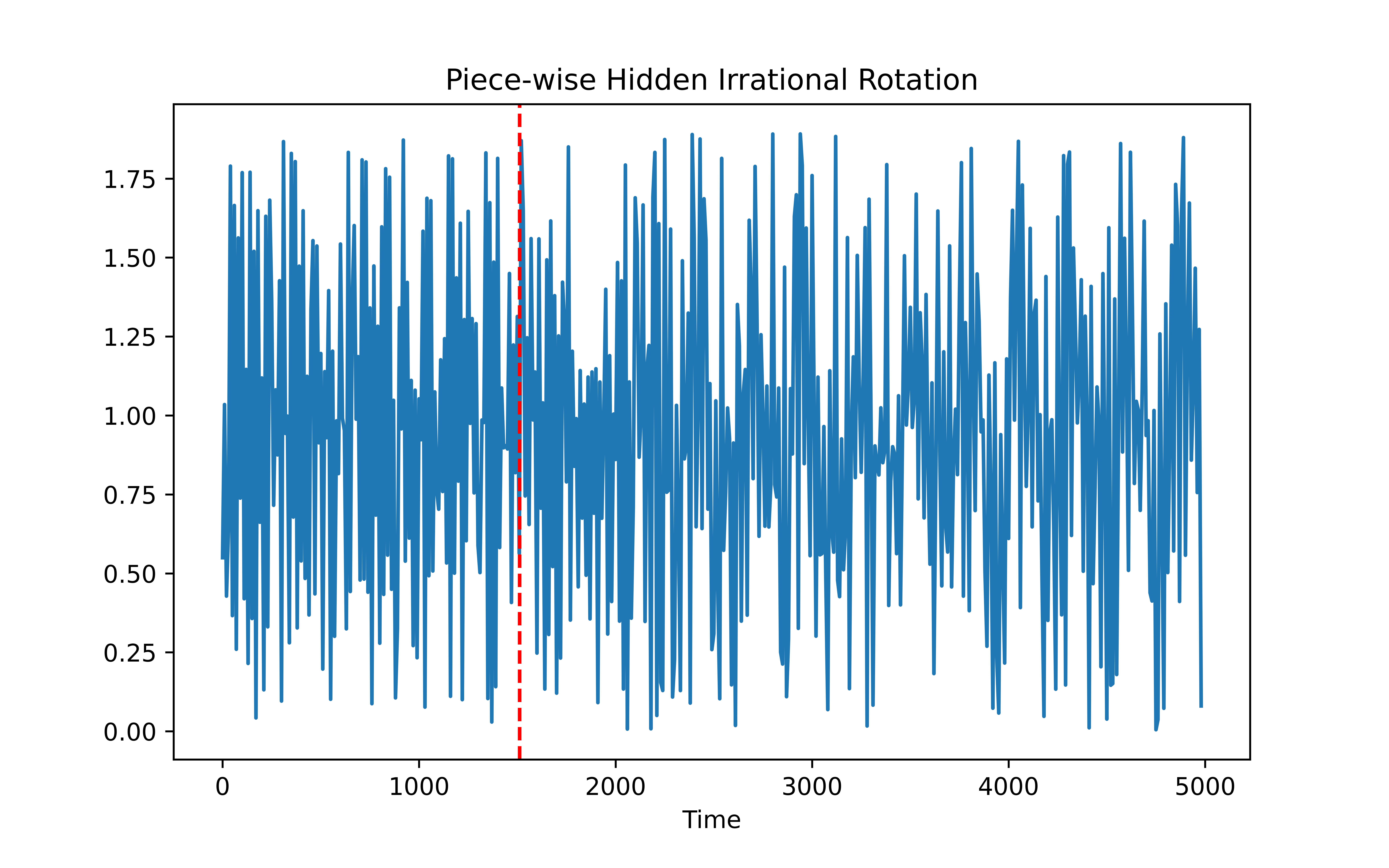}
\end{center}
\caption{An illustration of a piece-wise stationary sample generated by the concatenation of two different Hidden Irrational Rotations.   There is a changepoint at $1500$ and the segments before and after the change correspond to  sample-paths of different Hidden Irrational Rotations. }\label{fig:hir}
\end{figure}
In this section we compare the performance of \pkg{PyChEst} against that of \pkg{wbs}.  As mentioned previously,  the latter provides a consistent changepoint estimation method for the case where the samples are piece-wise i.i.d. .  

To calculate the performance of the algorithms,  we rely on their estimation errors  calculated as follows.  Suppose that there are a total of $\kappa$ changepoints $\tau_1(n),\dots,\tau(n)$  in a sample of length $n \in \N$.  We calculate the estimation error of an algorithm that produces a total of $K$ candidate estimates $\widehat{\tau}_1(n), \dots, \widehat{\tau}_{K}(n)$ as
\begin{align}\label{eq:error_rate}
e_n:= 
\begin{cases}
    \frac{1}{\kappa}\sum_i^{\kappa} |\tau_i(n) - \widehat{\tau}_i(n)|, & \text{if } K=\kappa\\
    1,              & \text{otherwise.}
\end{cases}
\end{align}

In our first experiment we examine the convergence of the estimation error $e_n$ as sample-length $n$ increases.  First,  we consider the problem of locating a single change in distribution so that $\kappa=1$.  We start with a simple setting where we generate piece-wise i.i.d.  samples of length $n = 500, 1000,  1500, 4500$,  each with a  changepoint at $\tau(n):= \lfloor 0.4 n\rfloor$; the samples before the changepoint are generated i.i.d.  according to a Bernoulli distribution with parameter $0.8$ and independently,  those after the changepoint are generated i.i.d.  according to a Bernoulli distribution with parameter $0.5$.  We set $\alpha:=0.21$ and provide \code{find\_changepoints}  with  \code{min_distance=0.21},  and \code{process\_count=2}.  We provide \pkg{wbs} with an upper-bound of $2$ on the number of changepoints together with the Strengthened Schwarz Information Criterion (SIC) penalty term as recommended by \citet{wbs-CRAN}.  As can be seen in Figure~\ref{fig:hir}(Left),  the mean estimation error calculated by \eqref{eq:error_rate} (averaged over $20$ iterations) converges to $0$ for both algorithms.  Next we repeated the same experiment,  but this time we let $\tau(n):= \lfloor 0.3 n\rfloor,~n =5000, 10000,15000, 20000,25000,30000$ and generated Hidden Irrational Rotations with shift parameters $0.452341643253462432$ and $0.6345354645623456234234$ before and after the changepoint respectively.  To run the algorithms,  we pass \code{min_distance=0.21},  and \code{process\_count=2} to \code{find\_changepoints},  and provided \pkg{wbs} with the same parameters as in the previous case,  namely an upper-bound of $4$ on the number of changepoints and the Strengthened SIC penalty term.  As shown in Figure~\ref{fig:hir} (Right),   the mean error rate (calculated over $20$ iterations) of \pkg{PyChEst} approaches $0$ with increased sample length $n$,  while that of \pkg{wbs} remains consistently high.  This experiment shows that the rate of convergence of $e_n$  for an algorithm such as \pkg{wbs} with consistency guarantees for the piece-wise i.i.d.  setting can be better than that of \pkg{PyChEst}'s when the i.i.d.  assumption holds.  However,  \pkg{PyChEst} is robust against long-range dependencies,  and performs well on both piece-wise i.i.d.  samples as well as piece-wise stationary samples. 
\begin{figure}
\begin{center}
\includegraphics[width=7.7cm]{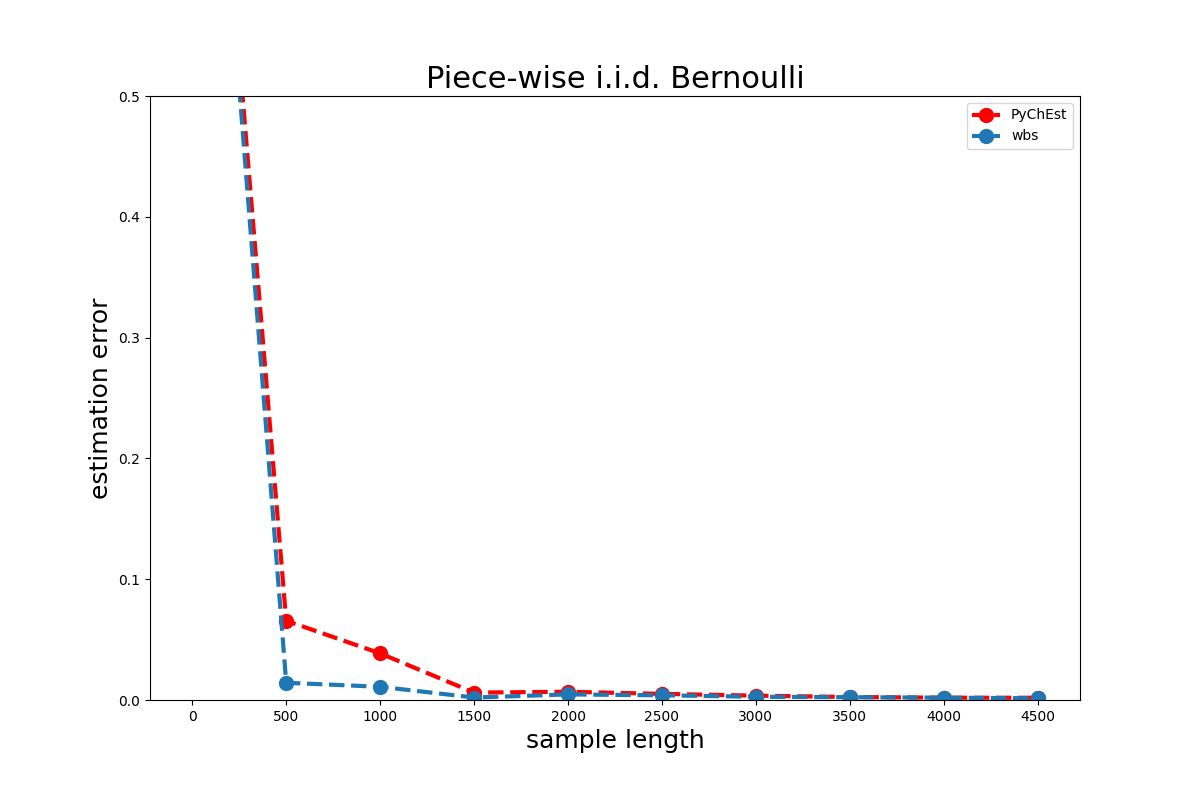}
\includegraphics[width=7.7cm]{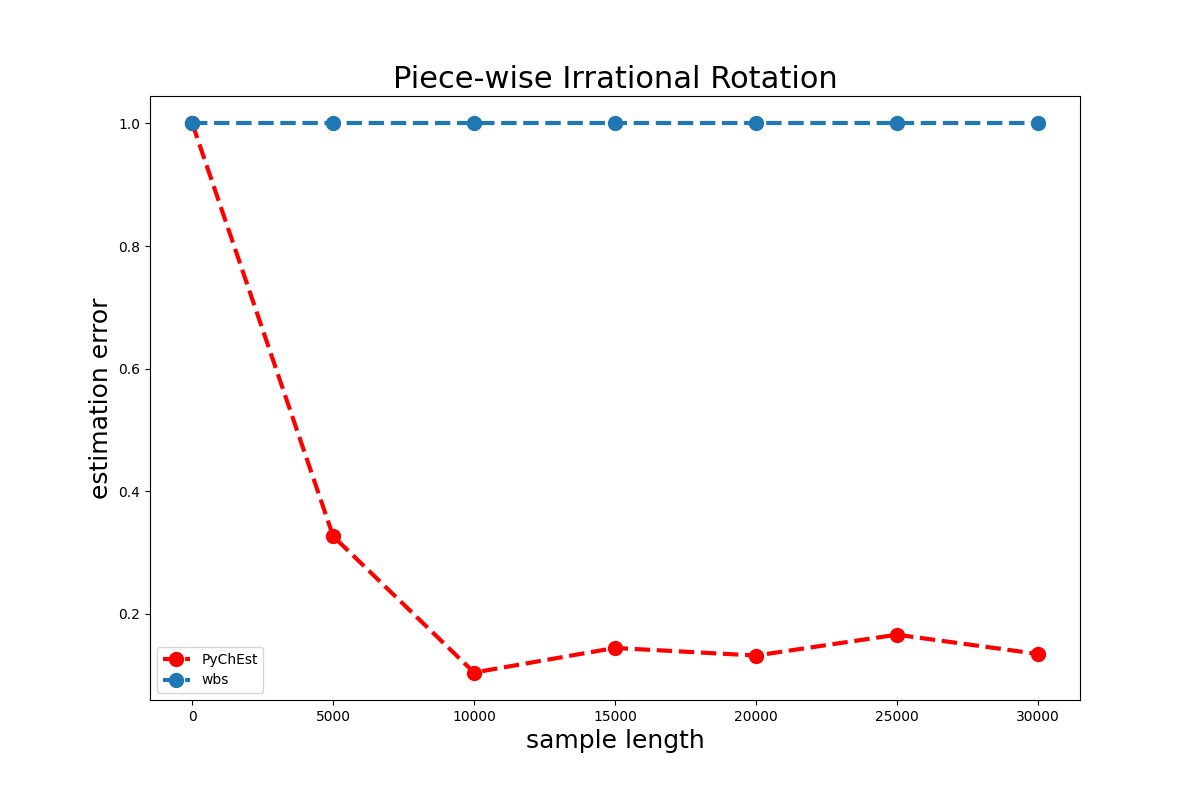}
\end{center}
\caption{Illustration of the robustness of \pkg{PyChEst} against long-range dependencies.  \textbf{Left:} Both \pkg{PyChEst} and \pkg{wbs} perform well on piece-i.i.d.  samples: the average estimation error \eqref{eq:error_rate} converges to $0$ for both algorithms.   \textbf{Right:} Despite long-range dependencies,  the error rate of \pkg{PyChEst} on locating a changepoint in a piece-wise stationary ergodic process obtained by a Hidden Irrational Rotation, converges with sample-length,  while that of \pkg{wbs} (which is provably consistent provided that the stationary samples are generated i.i.d.) remains consistently high in all $20$ iterations. }
\end{figure}

In our second experiment,  we consider the problem of locating $\kappa=2$ changepoints in piece-wise stationary samples of length $N=60000$.   First,  we generate piece-wise i.i.d.  Bernoulli samples with changepoints at $0.4N$ and $0.7N$,  so that $\lambda_{\min}=0.3$.  The three stationary segments are drawn independently where the first and last segments are generated i.i.d.  according to a Bernoulli distribution with parameter $0.8$ and the middle segment is generated i.i.d.  according to a Bernoulli distribution with parameter $0.5$,  leading to  $m=2$ as the total number of process distributions. 
We pass \code{process\_count=2} and \code{min\_distance=0.21} to \pkg{PyChEst}'s \code{find\_changepoints}.
We run \pkg{wbs} with an upper-bound of $4$ on $\kappa$ together with the Strengthened SIC penalty term.  Next,  we consider the same problem,  with the difference that the changepoints are located at $0.3N$ and $0.6N$ and the samples are piece-wise stationary ergodic processes with long-range dependencies.  More specifically,  the first and last segments are identically generated by a Hidden Irrational Rotation with shift parameter $0.452341643253462432$ and the middle segment is generated by a different Hidden Irrational Rotation with  $0.6345354645623456234234$  as its  shift parameter.  We run both algorithms with the same input parameters as those provided in the piece-wise i.i.d.  case above. 
The average (over $200$ iterations) errors~\eqref{eq:error_rate} is reported in Table~\ref{tab:exp_results}.  Both algorithms perform well in the piece-wise i.i.d.  setting.  However,  in the case of piece-wise Hidden Irrational Rotations  \pkg{wbs} is unable to provide a single changepoint estimate,  while \pkg{PyChEst} continues to perform well.  As a result,  this experiment serves as another demonstration of the robustness of \pkg{PyChEst} against long-range dependencies. 

\begin{table}
\begin{center}
\begin{tabular}{ |p{7cm}||p{3cm}|p{3cm}|  }
 \hline
 \multicolumn{3}{|c|}{piece-wise i.i.d.  vs.  piece-wise stationary process with long-range dependencies} \\
 \hline
& \pkg{PyChest} & \pkg{wbs} \\
 \hline
 (piece-wise i.i.d. ) Bernoulli           & 0.000151           & 0.000124                \\
 \hline
 (piece-wise) Hidden Irrational Rotation & 0.000143           & 1                      \\ 
 \hline
\end{tabular}
\caption{Average (over 200 iterations) error \eqref{eq:error_rate} on piece-wise stationary samples of length 60000.  In each case,  the samples have $\kappa=2$ changepoints where $\lambda_{\min} = 0.28$; each stationary segment is generated by one of $m=2$ different processes i.e.  the first and last processes are identical.  A  lower-bound $\alpha=0.21$ is provided as input to \code{find\_chanepoints} together with the correct number of processes $m=2$.   An upper-bound of $4$ on $\kappa$ together with the Strengthened SIC penalty term are passed to \pkg{wbs}.  Both algorithms perform well in the piece-wise i.i.d.  (Bernoulli) setting,  while \pkg{wbs} is consistently confused by the long-range dependencies in piece-wise Hidden Irrational Rotations.}
\label{tab:exp_results}
\end{center}
\end{table}

\end{document}